\documentclass[sort&compress,final,5p,twocolumn,times]{elsarticle}

\journal{Physics Letters B}

\usepackage{graphicx,bm,color}
\usepackage[dvipsnames]{xcolor}
\usepackage{amsmath}
\usepackage{amssymb}
\usepackage{booktabs}
\usepackage{float}
\usepackage[caption=false]{subfig} 
\usepackage{comment}

\usepackage[hidelinks]{hyperref}
 \hypersetup{
   pdftitle={},%
   pdfauthor={},%
   pdfsubject={},%
   pdfkeywords={},%
   pdfstartview={},
   bookmarksopen=true, breaklinks=true, debug=true, %
   colorlinks=true, linkcolor=red, citecolor=blue, urlcolor=blue
 }
\usepackage[normalem]{ulem}

\newcommand{\pjT}{p_{{\text j}T}}

\newcommand*\oline[1]{%
   \vbox{%
     \hrule height 0.5pt
     \kern0.4ex
     \hbox{%
       \kern-0.15em
       \ifmmode#1\else\ensuremath{#1}\fi
       \kern-0.15em
     }
   }
}

\usepackage[normalem]{ulem}

\usepackage{xspace}

\begin{document}

\title{Collins asymmetries for pion-in-jet production in polarized $\ell p$ collisions at the EIC}

\date{\today}

\author[add1,add2]{Umberto D'Alesio\corref{cor1}}
\ead{umberto.dalesio@ca.infn.it}

\author[add1,add2]{Carlo Flore}
\ead{carlo.flore@unica.it}
\cortext[cor1]{Corresponding author}

\author[add3]{Marco Zaccheddu}
\ead{zacch@jlab.org}

\address[add1]{Dipartimento di Fisica, Universit\`a di Cagliari, Cittadella Universitaria, I-09042 Monserrato (CA), Italy}
\address[add2]{INFN, Sezione di Cagliari, Cittadella Universitaria, I-09042 Monserrato (CA), Italy}
\address[add3]{Theory Center, Jefferson Lab, 12000 Jefferson Avenue, Newport News, Virginia 23606, USA}

\begin{abstract}
We study Collins azimuthal asymmetries for pion-in-jet production in polarized lepton-proton collisions, extending previous analyses of polarized $pp$ scattering to a complementary and theoretically simpler process.
We keep adopting a simplified transverse momentum dependent (TMD) approach, with a collinear configuration for the initial state, and employ the transversity and Collins fragmentation functions as extracted from semi-inclusive deep inelastic scattering and $e^+ e^-$ annihilation processes.
We then compute azimuthal asymmetries for the Electron-Ion Collider (EIC) kinematics, both within a leading order (LO) approach and by including quasireal photon exchange in the Weizs\"acker-Williams approximation.
Although this contribution is relevant  in the whole kinematical range explored, it does not spoil the dominance of quark-initiated channels, leaving only a marginal role to their gluon counterparts.
In this respect, Collins asymmetries in lepton-proton processes allow for a much clearer access to the transversity distribution, including its sea-quark component.
As we will argue, a comparison with future EIC data could represent a further step in testing the hypothesis of the universality of the Collins function as well as of the TMD factorization for this class of processes.
\end{abstract}

\begin{keyword}
Collins and transversity functions \sep Azimuthal Asymmetries \sep Universality \sep JLAB-THY-26-4697
\end{keyword}

\maketitle
\section{\label{sec:intro}Introduction}

The study of the three-dimensional structure of hadrons, the hadron tomography, is nowadays a fundamental pillar in understanding how the structure and properties of hadrons emerge from the dynamics of confined quarks and gluons.
Azimuthal and transverse-spin asymmetries represent a formidable tool to extract this information, allowing to investigate hadron/parton spin and transverse momentum correlations both in the nucleon structure as well as in the hadronization mechanism.

The so-called transverse momentum dependent (TMD) approach~\cite{Kotzinian:1994dv,Tangerman:1994eh,Tangerman:1995hw,Boer:1997nt,Boer:1997mf,Bacchetta:2006tn,Arnold:2008kf,Pitonyak:2013dsu}, allowing for the separation of the soft physics from the hard, perturbatively calculable parts, is the main and consolidated framework to deal with these observables.
The crucial condition to adopt this formalism is the presence of two ordered energy scales~\cite{Collins:1981uk,Collins:1984kg, Ji:2004wu, Collins:2011zzd,Echevarria:2012js}.
This is what happens in semi-inclusive deep inelastic scattering (SIDIS), Drell-Yan (DY) processes and the almost back-to-back hadron-pair production in $e^+e^-$ collisions.

Experimental investigations carried out over the last twenty years, and still ongoing, at various  facilities~\cite{Belle:2008fdv,COMPASS:2008isr,HERMES:2009lmz,JeffersonLabHallA:2011ayy,
Accardi:2012qut, COMPASS:2012dmt,BaBar:2013jdt,Dudek:2012vr, Aschenauer:2015eha, BESIII:2015fyw,
Adams:2018pwt,SeaQuest:2019hsx,  Aidala:2019pit, LHCspin:2025lvj,  Accardi:2023chb}, combined with increasingly accurate phenomenological analyses~\cite{Anselmino:2005ea,Anselmino:2007fs,Anselmino:2008sga,Anselmino:2013vqa,Anselmino:2015sxa,Boglione:2018dqd, DAlesio:2020vtw, Echevarria:2020hpy, Bury:2021sue, Bacchetta:2022awv, Bacchetta:2024qre,Bacchetta:2025ara, Moos:2025sal, Boglione:2024dal, Barry:2025glq,Kang:2026mod}, have by now provided a rather solid and robust three-dimensional picture of hadrons.

In a series of  papers~\cite{Yuan:2007nd,DAlesio:2010sag,DAlesio:2017bvu,Kang:2017btw,Kang:2017glf}, and more recently in Ref.~\cite{DAlesio:2025jmr}, the study of hadron-in-jet production\footnote{For an interesting analogy between the inclusive production of dihadrons, namely of two hadrons originating from the fragmentation of the same parton, and of a hadron inside a jet,
see Ref.~\cite{Bacchetta:2023njc}.} in $pp$ collisions has been proposed as a tool to further investigate the fundamental issues of TMD factorization and the universality of the Collins fragmentation function (FF)~\cite{Collins:1992kk}. This TMD-FF, describing the fragmentation of a transversely polarized quark into an unpolarized hadron, is indeed predicted to be the same in SIDIS and $e^+e^-$ annihilation processes~\cite{Metz:2002iz,Boer:2003cm,Collins:2004nx,Yuan:2009dw,Boer:2010ya}.
This is in contrast to the Sivers function~\cite{Sivers:1989cc,Sivers:1990fh}, expected to change in sign when moving from SIDIS to DY processes~\cite{Collins:2002kn,Brodsky:2002rv,Boer:2003cm}.

The presence of two ordered energy scales  in $pp\to{\rm{jet}}\, h \, X$ processes (the large transverse momentum of the jet and the small intrinsic transverse momentum of the hadron with respect to the jet direction) supports the conjecture of the validity of the TMD factorization also for this case.
In Ref.~\cite{DAlesio:2025jmr}, within a leading-order (LO) TMD approach and a collinear initial-state configuration, we have analyzed STAR data~\cite{STAR:2022hqg, Zhang:2024zuq,STAR:2025xyp} on the Collins  asymmetry, by employing the Collins function and the transversity distribution as extracted from azimuthal asymmetries in SIDIS and $e^+e^-$ processes.
The main outcomes of this study are the reasonable data-theory agreement, the absence of sizable factorization breaking effects together with the universality of the Collins FF, as well as the marginal role played by TMD evolution.
The corresponding analysis of the transverse polarization of $\Lambda$ hyperons within a jet produced in unpolarized $pp$ collisions led to  similar conclusions~\cite{DAlesio:2024ope}.

In this paper we extend this approach to the corresponding but simpler process $\ell p^\uparrow \to {\rm jet}\, \pi\, X$, where the final-state lepton is not observed. 
While the SIDIS-like case is extremely interesting by itself, and is indeed under investigation, 
the presence of a third independent scale (the virtuality of the exchanged boson) would require a more detailed and different treatment (see also, e.g., Ref.~\cite{Arratia:2020nxw} for a study of hadron-in-jet production in a back-to-back lepton–jet configuration), spoiling at the same time the strong similarity with the $pp$ case.  

Here we would like to emphasize how $\ell p$ processes could play a complementary role in testing the universality and TMD factorization issues, allowing, at the same time, for a more direct access to the transversity function. The reason is that the LO contribution involves the standard lepton-quark scattering, without mixing up with gluons that play a non negligible role in the denominator of the Collins asymmetry in $pp$ collisions.

In order to improve this treatment, without embarking on the complete next-to-leading-order calculation, we include here the contribution from quasireal photon exchange. This could be potentially relevant in the kinematical configuration where the lepton in the final state moves collinearly with the initial lepton. To this aim we employ the Weizs\"acker-Williams (WW) or equivalent photon approximation (EPA)~\cite{vonWeizsacker:1934nji,Williams:1934ad}, where the incoming lepton acts as a source of real photons (see also Refs.~\cite{Brodsky:1971ud,Terazawa:1973tb,Kniehl:1990iv}).
As we will show, this contribution is indeed sizable. On the other hand, even if it opens other partonic channels, like those initiated by a gluon, it does not spoil the dominance of quark initiated sub-processes.

We will compute Collins azimuthal asymmetries for the Electron-Ion Collider (EIC) kinematics~\cite{AbdulKhalek:2021gbh}, exploring different and complementary regions in transverse momenta and rapidities both of the jet and the final hadron. 
This will allow us to highlight the great potential of this study and the importance of future measurements of these asymmetries. 
For completeness, we also present some estimates for the unpolarized cross sections, to fully appreciate the role of the WW contribution.
We will then show how in some kinematic regions one can directly access the sea-quark component of the transversity, still almost unknown. In this respect, the investigation of pion-in-jet production in $\ell p$ processes complements the already important information gathered from the analysis in $pp$ collisions and it is therefore worth investigating at the EIC. 

The paper is organized as follows: in Section~\ref{sec:formalism} we present the theoretical calculation starting from the LO treatment and then moving to the contribution of the quasireal photon exchange. In Section~\ref{sec:results} we show our predictions for EIC kinematics, both for the unpolarized cross sections and the azimuthal asymmetries. In Section~\ref{sec:conclusions} we draw our conclusions.

\section{\label{sec:formalism} Formalism}

We compute the azimuthal asymmetry for an unpolarized hadron $h$ produced within a jet in single-polarized lepton-nucleon ($\ell N$) collisions,
\begin{equation}
\ell(p_\ell)\, N^\uparrow (P)\rightarrow {\rm jet}(p_{\rm{j}})\,  h(p_h)\,X\, ,
\end{equation}
where $p_\ell, P, p_{\rm j}, p_h$ are the momenta of the incoming lepton, the incoming nucleon, the jet, and the produced unpolarized hadron respectively.

Following Ref.~\cite{DAlesio:2025jmr} we will include transverse momentum effects only in the fragmentation mechanism within a leading order factorization framework, assuming a collinear initial state. This scheme is indeed extremely close to a TMD factorization picture where two measurable and ordered scales appear: the large transverse momentum of the jet and the small transverse momentum of the pion with respect to the jet.

The kinematics adopted is the following one: we work in the $\ell N$ center-of-mass (c.m.) frame, with $N$ moving along the \emph {positive} $z$ axis and the jet laying in the $xz$ plane.
The four-momenta (for massless hadrons and partons) are given by
\begin{equation}
\begin{aligned}
&P^\mu = \frac{\sqrt s}{2}(1,0,0,1), \quad p_\ell^\mu  = \frac{\sqrt s}{2}(1,0,0,-1),\\
&p_{\rm j}^\mu = E_{\rm j} (1, \sin\theta_{\rm j}, 0,\cos\theta_{\rm j}) = \pjT (\cosh\eta_{\rm j},1,0,\sinh\eta_{\rm j})\,,
\end{aligned}
\end{equation}
where $s$ is the c.m.~energy squared, $\eta_{\rm j}=-\log[\tan (\theta_{\rm j}/2)]$, is the jet pseudorapidity and $\pjT\equiv |\bm{p}_{{\rm j}T}|$ its transverse momentum.

The Collins contribution to the asymmetry for the $\ell N^\uparrow \to {\rm jet}\,h\, X$ process can be extracted by computing
the following azimuthal moment:
\begin{equation}
\begin{aligned}
& A_{UT}^{\sin(\phi_S-\phi^H_h)}(\bm{p}_{\rm j},z,p_{\perp \pi}) \\
= & 2\,\frac{\int\,d\phi_S \,d\phi_{h}^H\,\sin(\phi_S-\phi_h^H)\,
[d\sigma(\phi_S,\phi_h^H)-d\sigma(\phi_S+\pi,\phi_h^H)]} {\int\,d\phi_S\, d\phi_{h}^H\,[d\sigma(\phi_S,\phi_h^H)+d\sigma(\phi_S+\pi,\phi_h^H)]}\,,
\end{aligned}
\label{eq:an-col}
\end{equation}
where $d\sigma(\phi_S,\phi_h^H)$ stands for the  differential cross section
\begin{equation}
\frac{E_{\rm j}\,d\sigma^{\ell\, N(S,\phi_S)\to {\rm jet}\,h(\phi_h^H)\,X}}
{d^3\bm{p}_{\rm j}\,dz\,d^2\bm{p}_{\perp h}\,\,d\phi_S}\,.
\label{eq:dsig}
\end{equation}
Here $z$ is the pion longitudinal momentum fraction\footnote{Focusing on light mesons, we will identify the light-cone momentum fraction with the longitudinal momentum fraction.}, $\bm{p}_{\perp h}\, \equiv \bm{j}_T$ ($p_{\perp h}=|\bm{p}_{\perp h}|$) is the transverse momentum of the pion with respect to the parent parton
, $\phi_h^H$ is the azimuthal angle of the final-state hadron momentum, measured in the jet helicity frame, and $S$ is the (transverse) polarization vector of the initial nucleon beam, forming an angle $\phi_S$ with the jet production plane (the $xz$ plane) in the $\ell N$ c.m.~reference frame.
The azimuthal factor $\sin(\phi_S-\phi^H_h)$ in the numerator of Eq.~(\ref{eq:an-col}) singles out the Collins contribution. We recall once again that we apply this TMD scheme in the region where $j_T\ll p_{{\rm j}T}$.

We will consider both the LO contribution as well as the quasireal photon exchange, adopting the Weizs\"acker-Williams approximation.
Therefore we will have
\begin{equation}
\label{NAsym}
 d\sigma(\phi_S,\phi_h^H) =
 d\sigma_{\rm LO}(\phi_S,\phi_h^H) + d\sigma_{\rm WW}(\phi_S,\phi_h^H)\,.
\end{equation}
Here below we give all relevant expressions considering, separately, the LO and the WW contributions entering the numerator and the denominator of the azimuthal asymmetry.
Notice that in both cases the final parton in the elementary scattering process will be identified with the jet. We will comment on this below.

For the first case we have a single partonic channel, namely $q\ell\to q'\ell'$ ($q' \equiv$ jet), with $p_q = x P$,
\begin{eqnarray}
2\,d\sigma_{\rm LO}^{\rm unp}(\phi_h^H) &\equiv&  [d\sigma(\phi_S,\phi_h^H)+d\sigma(\phi_S+\pi,\phi_h^H)]_{\rm LO}\nonumber \\
& = &\sum_{q}\int \frac{dx}{16\pi^2xs}\, \delta(\hat s + \hat t +\hat u) \nonumber\\
& \times &
f_{q/N}(x)\,(|\hat M_1|^2+|\hat M_2|^2)_{q\ell}\,D_{h/q}(z,p_{\perp h}^2)\,, \nonumber\\
\label{DenLO}
\end{eqnarray}
\begin{eqnarray}
d\Delta\sigma_{\rm LO} &\equiv& [d\sigma(\phi_S,\phi_h^H)-d\sigma(\phi_S+\pi,\phi_h^H)]_{\rm LO} \nonumber\\
& = &\sin(\phi_S-\phi_h^H)\, \sum_{q}\int\frac{dx}{16\pi^2xs}\,\delta(\hat s + \hat t +\hat u)\nonumber\\
&\times  &
h_{1}^q(x)\,(\hat M_1\hat M_2)_{q\ell} \, \Delta^N\! D_{h/q^\uparrow}(z,p_{\perp h}^2)\,, \label{NumLO}
\end{eqnarray}
where $\hat M_i$'s ($i=1,2$) are the two independent helicity amplitudes for the elementary scattering process $q\ell\to q'\ell'$ (see, e.g., Ref.~\cite{Anselmino:2009pn}, where they are indicated as $\hat M_i^0$):
\begin{eqnarray}
(|\hat M_1|^2+|\hat M_2|^2)_{q\ell} & = & 4\, e_q^2\, e^4 \,
\Big(\frac{\hat s^2+\hat u^2}{\hat t^2} \Big) \label{ampl2} \\
(\hat M_1 \, \hat M_2)_{q\ell} & = & 4\, e_q^2\, e^4\,
 \Big(-\frac{\hat s\hat u}{\hat t^2}\Big) \label{ampl} \>.
\end{eqnarray}
In the following we use $e^2 = 4\pi\alpha$, $\alpha$ being the electromagnetic coupling constant.
The Mandelstam invariants for the elementary process $q(p_q)\,\ell(p_\ell)\to q'(p_q') \,\ell' (p_\ell')$ are
\begin{equation}
\hat s = (p_q+p_\ell)^2 \quad \hat t= (p_q-p_q')^2 \quad \hat u = (p_\ell - p_q')^2   \,,
\end{equation}
that, with $p'_q = p_{\rm j}$, can be expressed as:
\begin{equation}
\hat s = x s\, \quad \hat t = -Q^2=-x\sqrt s\, p_{{\rm j}T}\,e^{-\eta_{\rm j}}  \quad     \hat u =-\sqrt s\, p_{{\rm j}T}\,e^{\eta_{\rm j}}\,.
\end{equation}

Notice that the $\delta (\hat s+ \hat t + \hat u)$ in Eqs.~(\ref{DenLO}) and (\ref{NumLO}) allows us to carry out the integration over $x$ fixing it at
\begin{equation}
x= \frac{x_T e^{\eta_{\rm j}}}{2-x_Te^{-\eta_{\rm j}}}\,,
\end{equation}
with $x_T = 2 p_{{\rm j}T}/\sqrt s$.
It is easy to check that this value coincides with the standard Bjorken variable, $x_{\rm Bj} = Q^2/2q\cdot P$, here given in terms of the final quark variables (the jet in our LO treatment).

As one can see, the numerator in Eq.~(\ref{NumLO}) involves a convolution of the {\em collinear} quark transversity distribution, $h_1^q$, with the Collins TMD-FF, $\Delta^N\! D_{h/q^\uparrow}$, while the denominator, Eq.~(\ref{DenLO}), is simply twice the unpolarized cross section. The scale dependence of the nonperturbative functions in the above equations has been understood.

Concerning the WW piece we recall that for the inclusive process $\ell N\to h\,X$, at NLO the collinear lepton singularities could be regularized by introducing a QED parton distribution for the lepton. Similarly, for the process under study, we can say that at ${\cal O}(\alpha^2\alpha_s$) the leading term at $Q^2\simeq 0$ comes from a collinear photon acting as a parton of the lepton and entering the hard scattering process. This can be expressed as a Weizs\"acker-Williams contribution, where the lepton acts as a source of real photons. We then assume the following factorization formula for the WW contribution to the process $\ell N \to {\rm jet}\, h\, X$:
\begin{equation}
\label{WW}
\sigma_{\rm WW}(\ell N \to {\rm jet}\, h\, X) = \int\! d y \, f_{\gamma/\ell}(y)\, \sigma(\gamma  N\to {\rm jet} \, h\, X) \,,
\end{equation}
where $f_{\gamma/\ell}(y)$ is the distribution of photons inside the lepton, carrying a lepton-momentum fraction $y$ ($p_\gamma = y p_\ell$) and $\sigma(\gamma  N \to {\rm jet}\, h \,X)$ is the cross section for the process $\gamma N \to {\rm jet}\, h \,X$ initiated by a real photon.

Following Ref.~\cite{Kniehl:1996we}, 
for the WW distribution we adopt
\begin{equation}
\label{ww}
\begin{aligned}
 &f_{\gamma/\ell}(y) = \frac{\alpha}{2\pi}
\\& \times \Biggl[\frac{1 + (1-y)^2}{y}\ln\frac{Q^2_{\rm max}}{Q^2_{\rm min}(y)}+ 2m_\ell^2 y\left(\frac{1}{Q^2_{\rm max}} - \frac{1}{Q^2_{\rm min}(y)}\right) \Biggr]\,,\\[1mm]
\end{aligned}
\end{equation}
where $m_\ell$ is the lepton mass, $Q^2_{\rm max}~=1$~GeV$^2$ and $Q^2_{\rm min}(y)=m_\ell^2 y^2 / (1-y)$.

We are now equipped to give the expressions for the WW contributions. For the denominator in such a case we have to consider the following partonic channels: $q\gamma\to q g$, $q\gamma \to g q$ and $g\gamma\to q\bar q$, with $p_a = x P$  ($a=q,g$), giving
\begin{eqnarray}
 2d\sigma_{\rm WW}^{\rm unp}(\phi_h^H) &\equiv & [d\sigma(\phi_S,\phi_h^H)+d\sigma(\phi_S+\pi,\phi_h^H)]_{\rm WW}\nonumber \\
&=&  \sum_{a,c,d}\int\! dy\frac{dx}{16\pi^2\hat s}\, \delta(\hat s + \hat t +\hat u)
\,2\,\Sigma^{\rm unp}_{a\gamma\to cd}\,,\nonumber\\
\label{DenWW}
\end{eqnarray}
where
\begin{eqnarray}
2\,\Sigma^{\rm unp}_{q\gamma\to qg} &=& f_{\gamma/\ell}(y)
f_{q/N}(x)\,(|\hat M_1|^2+|\hat M_2|^2)_{q\gamma}\, D_{h/q}(z,p_{\perp h}^2)\nonumber\\
2\,\Sigma^{\rm unp}_{q\gamma\to gq} &=& f_{\gamma/\ell}(y)
f_{q/N}(x)\,(|\hat M_1|^2+|\hat M_3|^2)_{q\gamma}\, D_{h/g}(z,p_{\perp h}^2)\nonumber\\
2\,\Sigma^{\rm unp}_{g\gamma\to q\bar q} &=& f_{\gamma/\ell}(y)
f_{g/N}(x)\,(|\hat M_2|^2+|\hat M_3|^2)_{g\gamma}\, D_{h/q}(z,p_{\perp h}^2)\,,\nonumber\\
\label{eq:WWKernel}
\end{eqnarray}
with
\begin{eqnarray}
(|\hat M_1|^2+|\hat M_2|^2)_{q\gamma} & = & \frac{16}{3} \, e_q^2\, e^2\, g_s^2\,\Big[-\Big(\frac{\hat s}{\hat u}+\frac{\hat u}{\hat s} \Big)\Big] \\
(|\hat M_1|^2+|\hat M_3|^2)_{q\gamma} & = & \frac{16}{3} \, e_q^2\, e^2\, g_s^2\, \Big[-\Big(\frac{\hat s}{\hat t}+\frac{\hat t}{\hat s} \Big)\Big]\\
(|\hat M_2|^2+|\hat M_3|^2)_{g\gamma} & = & 2 \, e_q^2\, e^2\, g_s^2\, \Big(\frac{\hat u}{\hat t    }+\frac{\hat t}{\hat u} \Big)\,,
\end{eqnarray}
with $g_s^2 = 4 \pi \alpha_s$.

Notice that the three independent helicity amplitudes are different from the previous ones entering the LO contribution, see Ref.~\cite{DAlesio:2017nrd} for further details.
In this case, the Mandelstam variables for the elementary process $a(p_a)\,\gamma(p_\gamma) \to c(p_c) \, d(p_d)$ are
\begin{equation}
\hat s = (p_a+p_\gamma)^2 \quad \hat t= (p_a-p_c)^2 \quad \hat u = (p_\gamma - p_c)^2   \,,
\end{equation}
 and $p_c= p_{\rm j}$. More explicitly we have
\begin{equation}
\hat s = x ys\, \quad \hat t =-x\sqrt s\, p_{{\rm j}T}\,e^{-\eta_{\rm j}}  \quad     \hat u =-y\sqrt s\, p_{{\rm j}T}\,e^{\eta_{\rm j}}\,.
\end{equation}

The partonic channel $g\gamma\to \bar q q$, to be included as well, can be simply obtained from the channel $g\gamma\to q \bar q$ by interchanging in the corresponding equations $\hat u\leftrightarrow \hat t $ and replacing $D_{h/q}$ with $D_{h/\bar q}$.

In contrast, for the numerator of the azimuthal asymmetry only one channel contributes\footnote{All other spin transfers are zero (see Ref.~\cite{DAlesio:2017nrd}) and no gluon transversity exists for a spin-1/2 target.}, namely $q\gamma\to q g$, and we obtain:
\begin{eqnarray}
d\Delta\sigma_{\rm WW}&\equiv& [d\sigma(\phi_S,\phi_h^H)-d\sigma(\phi_S+\pi,\phi_h^H)]_{\rm WW} \nonumber\\
&=& \sin(\phi_S-\phi_h^H)\, \sum_q\int\! dy\frac{dx}{16\pi^2\hat s}\,\delta(\hat s + \hat t +\hat u)\nonumber\\
&\times&
f_{\gamma/\ell}(y)\,h_1^q(x)\,(\hat M_1\hat M_2)_{q\gamma} \, \Delta^N\! D_{h/q^\uparrow}(z,p_{\perp h}^2)\,,
\label{NumWW}
\end{eqnarray}
where \cite{DAlesio:2017nrd}
\begin{eqnarray}
(\hat M_1 \hat M_2)_{q\gamma} & = & \frac{16}{3}\, e_q^2\,  e^2\, g_s^2  \>.
\end{eqnarray}
Once again the numerator involves a convolution of the transversity and the Collins FF.

By exploiting the delta-function in Eqs.~\eqref{DenWW} and \eqref{NumWW}, one can integrate, for instance, over the variable $y$, so that
\begin{eqnarray}
d\sigma_{\rm WW}^{\rm unp}
=
\sum_{a,c,d}\int \!\frac{dx}{16\pi^2\hat s}\,
\frac{1}{x s - \sqrt s \,p_{{\rm j}T}\,e^{\eta_{\rm j}}}
\,\Sigma^{\rm unp}_{a\gamma\to cd}
\label{DenWW2}
\end{eqnarray}
\begin{eqnarray}
d\Delta\sigma_{\rm WW} & = &
\sin(\phi_S-\phi_h^H)
\sum_q\!\int\! \frac{dx}{16\pi^2\hat s}\frac{1}{x s - \sqrt s  \, p_{{\rm j}T}\,e^{\eta_{\rm j}}} \nonumber\\
&\times&
f_{\gamma/\ell}(y)\,h_1^q(x)\,(\hat M_1\hat M_2)_{q\gamma} \, \Delta^N\! D_{h/q^\uparrow}(z,p_{\perp h}^2),
\label{NumWW2}
\end{eqnarray}
with
\begin{equation}
y = \frac{x \, p_{{\rm j}T}\,e^{-\eta_{\rm j}}}{x \sqrt s - p_{{\rm j}T}\,e^{\eta_{\rm j}}}\,.
\end{equation}
By imposing $y\le 1$ we get the minimum value of $x$, being
\begin{equation}
x_{\rm min}= \frac{x_T e^{\eta_{\rm j}}}{2-x_Te^{-\eta_{\rm j}}}\,.
\end{equation}
Concerning the parameterizations for the transversity and the Collins FFs we follow Refs.~\cite{Anselmino:2007fs, Anselmino:2013vqa, Anselmino:2015sxa, DAlesio:2020vtw,DAlesio:2025jmr}. Namely, we use:
\begin{equation}\label{eq:h1(x)-SB}
 h_1^q(x\,, Q_0^2) = {\mathcal N}^T_q(x) \frac12 \left[f_{q/p}(x\,, Q_0^2) + g_{1L}^q(x\,, Q_0^2)\right]\,,
\end{equation}
with $Q_0^2 = 0.81\,\text{GeV}^2$ and where
\begin{equation}
 {\cal N}^{T}_q(x)=N^{T}_q x^{\alpha}(1-x)^\beta\,
\frac{(\alpha+\beta)^{\alpha+\beta}}{\alpha^\alpha \beta^\beta},
\quad (q = u_v,\,d_v)\,,
\end{equation}
with $\alpha$ and $\beta$ as extracted in  Ref.~\cite{Boglione:2024dal} and considering only valence contributions. 
For the Collins functions we have~\cite{Bacchetta:2004jz}:
\begin{eqnarray}
\label{eq:Collins}
\Delta^N\! D_{h/q^\uparrow}(z,p_{\perp h}^2) & = & \frac{2p_{\perp h}}{zm_h}\, H_1^{\perp q}(z, p_{\perp h}^2)\\
H_1^{\perp q}(z, p_{\perp h}^2)  &=& {\cal N}^C_q (z) \,\frac{z m_h}{ M_C}\,\sqrt{2e}\,e^{-p_\perp^2/M_C^2}\,D_{h/q}(z, p_{\perp h}^2)\,,\nonumber
\end{eqnarray}
where $m_h$ is the produced hadron mass and $D_{h/q}(z, p_{\perp h}^2)$ is the unpolarized TMD-FF, parameterized as:
\begin{equation}
\begin{aligned}
 & D_{h/q}(z, p_{\perp h}^2) = D_{h/q}(z)\, \frac{e^{-p^2_{\perp h} / \langle p^2_{\perp h} \rangle}}{\pi \langle p^2_{\perp h}\rangle}\,.
 \label{eq:unpTMD}
\end{aligned}
\end{equation}

In the following we focus on pion production ($h\equiv\pi$) and use $q = \rm{fav}, \rm{unf}$ (favored/unfavored FFs) in Eq.~(\ref{eq:Collins}), with the ${\cal N}^C_q(z)$ factors~\cite{Boglione:2024dal} given by
\begin{equation}
\label{eq:Collins-NC}
 {\cal N}^C_{\rm{fav}}(z) = N^C_{\rm{fav}}\, z^\gamma,
 \quad{\cal N}^C_{\rm{unf}}(z) = N^C_{\rm{unf}}\,,
\end{equation}
and $\langle p_{\perp\pi}^2\rangle = 0.12$ GeV$^2$ in Eq.~(\ref{eq:unpTMD}) as extracted in  Ref.~\cite{Anselmino:2013lza}\footnote{Notice that for the corresponding, still unknown, unpolarized gluon TMD-FF entering Eq.~(\ref{eq:WWKernel}) we use the same functional form, with the same Gaussian width, as for quarks.}.

We recall once again that these parametrizations of the  transversity and the Collins FFs have been extracted by fitting data on SIDIS and $e^+e^-$ azimuthal asymmetries.

For the collinear distributions we use the MSHT20nlo proton PDF set~\cite{Bailey:2020ooq} and the DEHSS set for pion fragmentation functions~\cite{deFlorian:2014xna,deFlorian:2017lwf}.
The transversity function in Eq.~\eqref{eq:h1(x)-SB} is then evolved to higher $Q^2$ values through a modified version of {\tt HOPPET}~\cite{Salam:2008qg,Prokudin:hoppet}, while in all other cases, including the collinear part of the Collins function (see Eq.~(\ref{eq:Collins})), we adopt DGLAP evolution.

A comment on the choice of the factorization scale, both in the collinear PDFs and in the TMD-FFs, is mandatory. On the one hand, the relevant scale for the TMD-FFs for this kind of processes is $\mu_j=\pjT R$, where $R$ is the jet-cone radius, as discussed in Ref.~\cite{Kang:2017glf}. Then, by evolving up to $\mu=p_{{\rm j}T}$, one resums single logarithms in the jet size parameter to all orders in the strong coupling constant $\alpha_s$. On the other hand, working here at LO accuracy, which is independent from $R$ (and more generally from the jet dynamics), and following
Ref.~\cite{Kang:2017btw}, we can safely use $\mu=\pjT$ for the unpolarized and the Collins FFs and similarly for the collinear distributions. This aspect deserves indeed more care and will be addressed in a future study, where we plan go beyond the LO accuracy and suitably employ the TMD jet fragmentation functions (TMDjFFs), see, e.g.,~Refs.~\cite{Kang:2017glf, Kang:2020xyq, Kang:2023elg}.

It is also worth mentioning that, in order to determine the proper partonic frame, one should consider the backward going jet together with the jet around which we measure the final hadron, as pointed out in Ref.~\cite{Boer:2007nh}. Indeed, the $z$ and $p_\perp$ dependencies of the FF for a jet with a large transverse momentum are not invariant under Lorentz boosts along the $\ell p$ direction. This could cause a mismatch with the corresponding variables adopted in SIDIS and $e^+e^-$ expressions. Since these differences scale as $E_{\rm j}/\sqrt s$~\cite{Boer:2010yp}, we expect that, for the energies and the kinematics considered here, these corrections are not larger than a few percent.

\begin{figure*}[t]
\centering
\includegraphics[width=\textwidth,keepaspectratio]{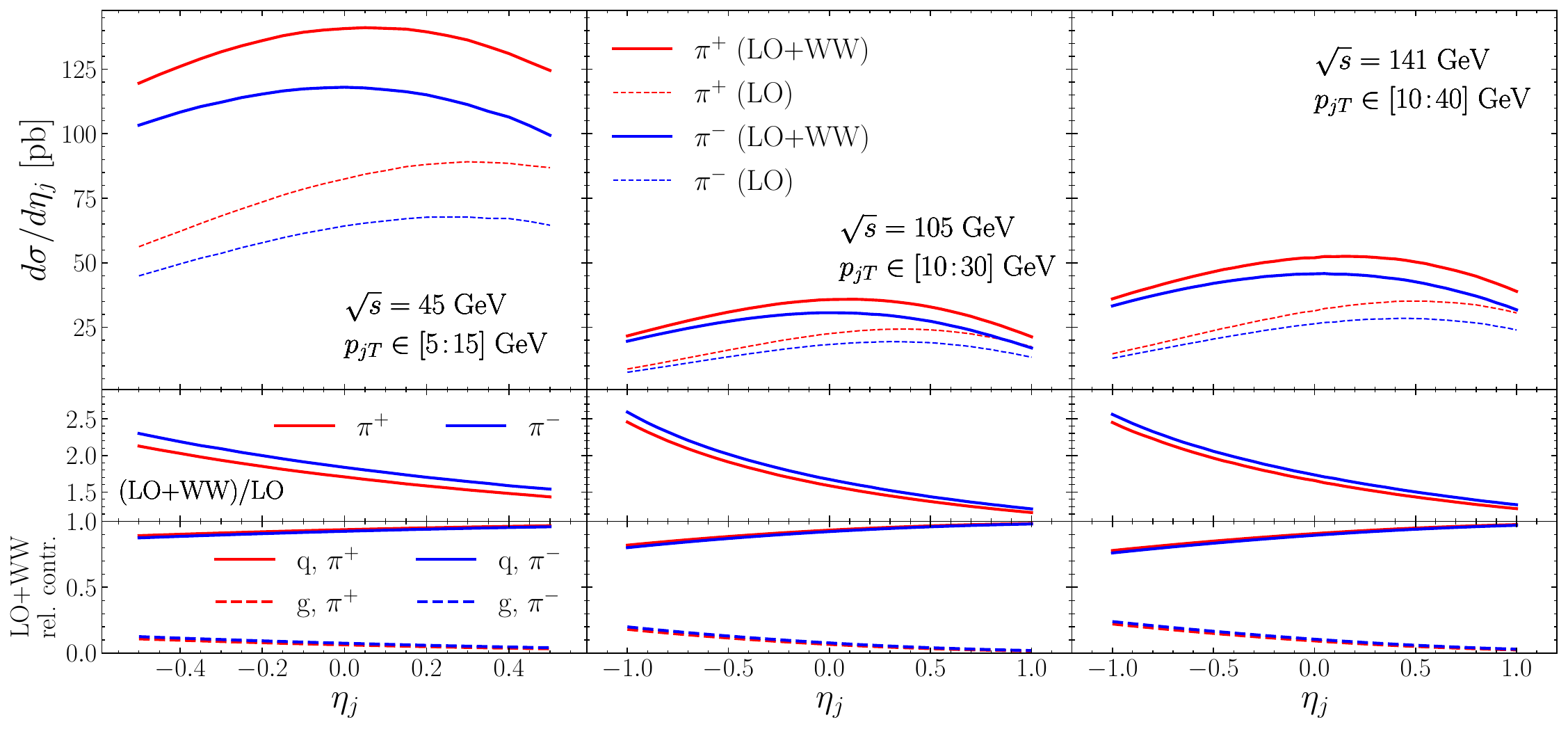}
\caption{Unpolarized cross sections for $\ell p \to {\rm jet}\,\pi^\pm\,X$ as a function of $\eta_{\rm j}$ at three different values of $\sqrt{s}$, with the corresponding $p_{{\rm j}T}$ ranges, see legend. Upper panels: LO and (LO+WW) results;
middle panels: ratio of full (LO+WW) contribution to the LO piece; lower panels: relative contribution of quark- and gluon-induced channels to the total cross section.}
\label{fig:dsig-detaj}
\end{figure*}

\section{\label{sec:results} Results}
We present here our theoretical estimates of the Collins azimuthal asymmetry for $\ell p^\uparrow \to {\rm jet}\, \pi\, X$ (that is for a proton target), for charged pions and focusing on EIC kinematics. More precisely, we consider three different center-of-mass energy values, $\sqrt s = $ 45, 105 and 141 GeV, forward and backward jet-production regions\footnote{We always refer to the proton direction, if not otherwise specified.}, with $j_T$ (the transverse momentum of the pion with respect to the jet) in the range 0.1-1.5 GeV, $z$ (the pion's light-cone momentum fraction) in the range 0.1-0.8 and the jet-cone radius $R$ between 0.05 an 0.6. Concerning the transverse momentum of the pion, $p_{{\rm j}T}$, the hard scale in the process, we consider the following ranges:
[5:15], [10:30] and [10:40] GeV respectively for the three energies above, allowing to explore comparable ranges in the $x_T= 2 p_{{\rm j}T}/\sqrt s$ variable. This will play a role in the phenomenological discussion below.
We also recall that when we show estimates as a function of one variable it is understood that we integrate over the other ones within the  ranges specified above.
While these values may not be necessarily the ones effectively explored at the EIC, they are representative for the study we are carrying out.

As already mentioned, in the process under consideration the final lepton is not observed.
By taking the transverse momentum of the jet, the final-state quark in our case, large enough, we ensure that at LO the partonic process is hard: $Q^2=-\hat t$ is large.
On the other hand, in order to include contributions where the final lepton goes in the forward direction (w.r.t.~the initial lepton direction) we will add the quasireal photon exchange by employing the WW approximation.

Before discussing the results on the azimuthal asymmetries, it is worth considering, and quantifying, the role played by the WW term in the unpolarized cross sections, focusing in particular on the quark-initiated channel.
In the following, for the sake of space, we will concentrate  only on the $\eta_{\rm j}$ dependence, for which we give our estimates.

Concerning the first issue,  it turns out that the WW piece contributes heavily, see Fig.~\ref{fig:dsig-detaj} (upper panels). In particular, the ratio of the \emph{full} result (LO+WW) to the LO contribution (middle panels) is at least 2 in the backward region and still larger than 1.5 in the forward region . The main reason of this different behavior is that in the backward region the gluon-initiated channels, entering only via the WW term, play a slightly more significant role.

Regarding the relative contribution of the quark-initiated channel to the full result, still focusing on the $\eta_{\rm j}$ distribution,  it turns out to be at least 80-90\%, depending on the c.m.~energy (Fig.~\ref{fig:dsig-detaj}, lower panels).
Similar behavior has been found for the $x_T$, $j_T$ and $z$ dependencies.

This is an important aspect since in the numerator of the azimuthal asymmetry only quark-initiated channels enter, driven by the quark transversity distribution. In this respect the marginal role played by the gluon contribution in the denominator (even when adding the WW piece) allows for a clear access to this distribution. This is at variance w.r.t.~the azimuthal asymmetry in $pp$ collisions, where the denominator is strongly affected/dominated by the gluon contribution.

\begin{figure*}[h!]
\centering
\includegraphics[width=17cm, keepaspectratio]{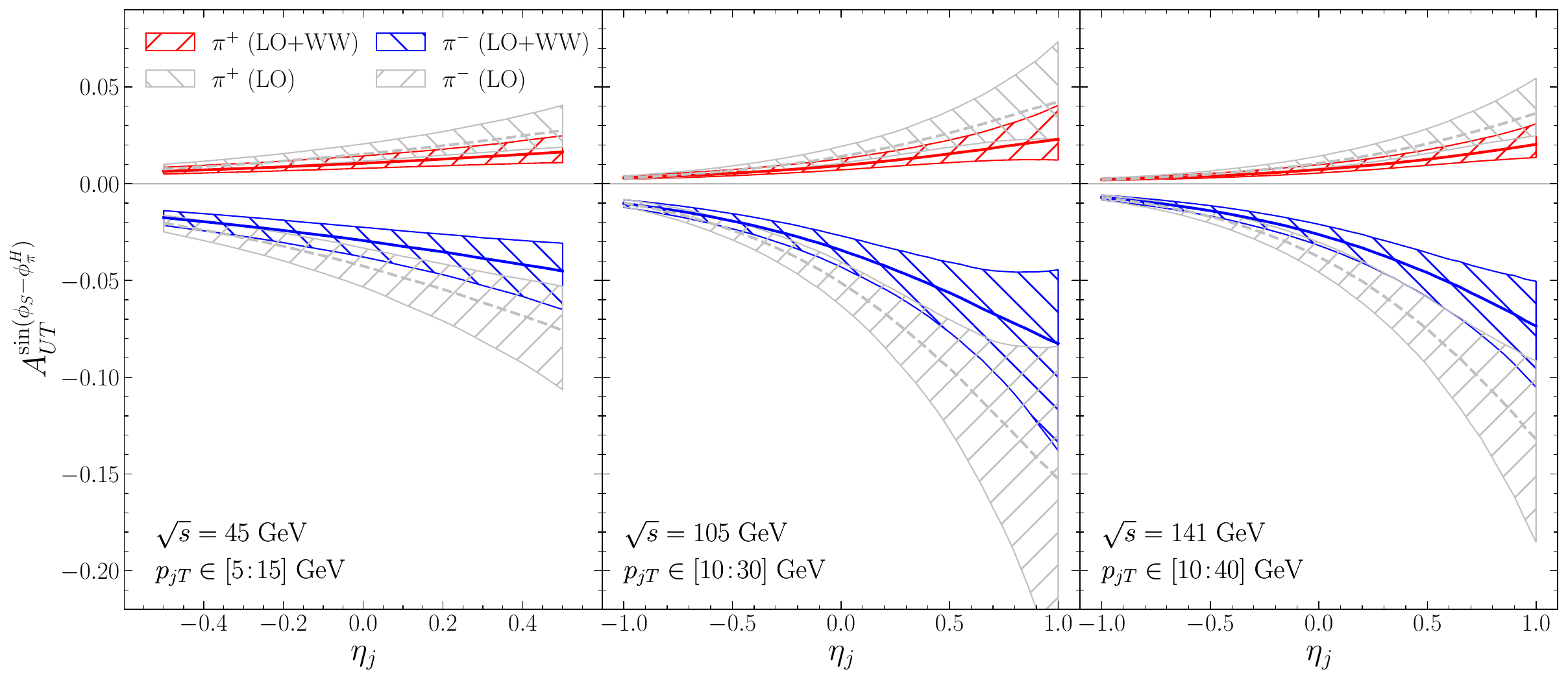}
\caption{Collins asymmetries for $\ell p^\uparrow \to {\rm jet}\,\pi^\pm\,X$ as a function of $\eta_{\rm j}$; LO (dashed lines and gray bands) and LO+WW (solid lines and colored bands). Left to right: $\sqrt s=$ 45, 105, 141 GeV, with the corresponding $p_{{\rm j}T}$ ranges, see legend. Uncertainty bands at $2\sigma$ CL.}
\label{fig:AUT-vs-etaj}
\end{figure*}

Let us now move to the azimuthal asymmetries, starting by looking at their $\eta_j$ dependence.
In Fig.~\ref{fig:AUT-vs-etaj} we present theoretical estimates for the three c.m.~energies together with the corresponding uncertainty bands. Here and in the following, we adopt the median of the Monte Carlo sets (generated to estimate the uncertainty on the extraction) as central value, and the uncertainties are computed at a 2$\sigma$ confidence level (CL).

As a general trend, the asymmetry tends to decrease in size from the LO to the LO+WW case, reflecting the increase of the unpolarized cross section (see Fig.~\ref{fig:dsig-detaj}). Moreover, while for $\pi^+$ production the asymmetries are relatively small, around $1\%$, the corresponding ones for $\pi^-$ tend to grow at forward rapidity, ranging from 2 to 4-8\%, depending on the $\sqrt{s}$ value.
This behavior can be understood by taking into account the following aspects: $i)$ since the integration is performed over the full $z$ range, the contribution from the favored Collins fragmentation function is partially suppressed with respect to that arising from the unfavored one (see Eq.~(\ref{eq:Collins-NC})); $ii)$ moving to increasing positive rapidities, that implies probing larger values of $x$, the role played by the up-quark transversity distribution (coupled to the unfavored/favored Collins FF for $\pi^-/\pi^+$ production) becomes more relevant.

It is also worth emphasizing that at large negative pseudorapidities (even larger than those considered here) one probes relatively small-$x$ values (of order $10^{-2}$), an effect that becomes increasingly relevant at higher c.m.~energies. Consequently, when a purely valence‑like parametrization of the transversity distribution is employed, as in the present case, the predicted asymmetries are markedly suppressed. This observation indicates that this kinematic region may provide a particularly sensitive probe of the sea‑quark component of transversity.

In the case of the other kinematical dependencies ($x_T$, $z$ and $j_T$), the separation between forward and backward rapidities requires more care as we will discuss below.

\begin{figure*}[t]
\centering
\includegraphics[width=16cm, keepaspectratio]{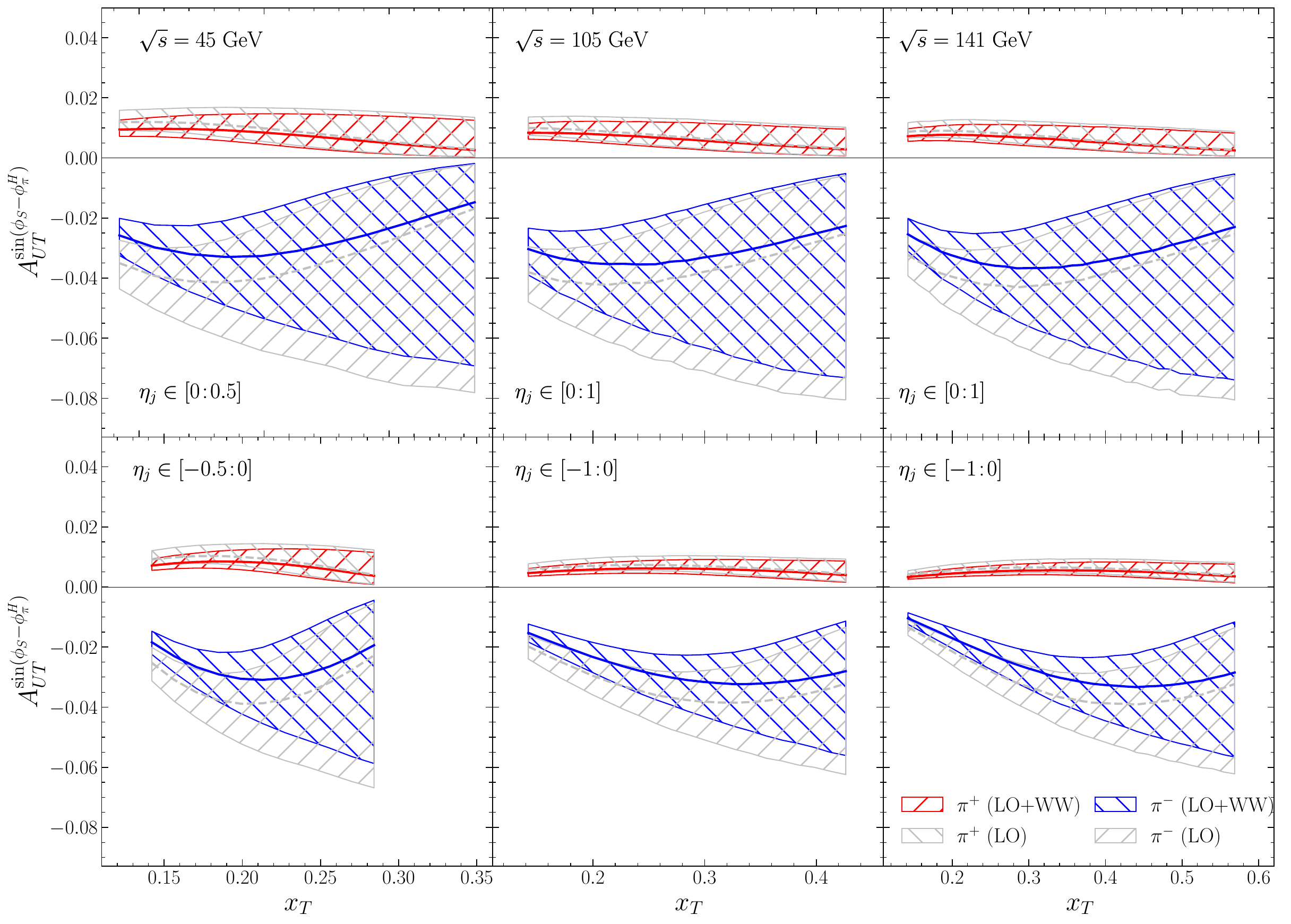}
\caption{Collins asymmetries for $\ell p^\uparrow\to {\rm jet}\,\pi^\pm\,X$ as a function of $x_T$: LO (dashed lines and gray bands) and LO+WW (solid lines and colored bands). Left to right: $\sqrt s=$ 45, 105, 141 GeV. Upper panels: forward rapidities; lower panels: backward rapidities. Uncertainty bands at $2\sigma$ CL.}
\label{fig:AUT-vs-xT}
\end{figure*}

In Fig.~\ref{fig:AUT-vs-xT} we show the azimuthal asymmetries as a function of $x_T$ for the three c.m.~energies.
As one can see, focusing on the [LO+WW] results, for $\pi^+$ we get estimates around 1\%, while for $\pi^-$  we reach larger values, in size: around 3-4\% in the forward region and 2-3\% in the backward region. This last behavior is understandable in terms of the increasing role of the gluons in the backward region. Moreover, similarly to the arguments mentioned above, the relatively large size of the unfavored Collins FF, see Eq.~(\ref{eq:Collins-NC}), when integrated over $z$ from the lowest value considered, coupled to the sizeable transversity for up quarks explains the larger effect for $\pi^-$ production.

\begin{figure*}[h!]
\centering
\includegraphics[width=16cm, keepaspectratio]{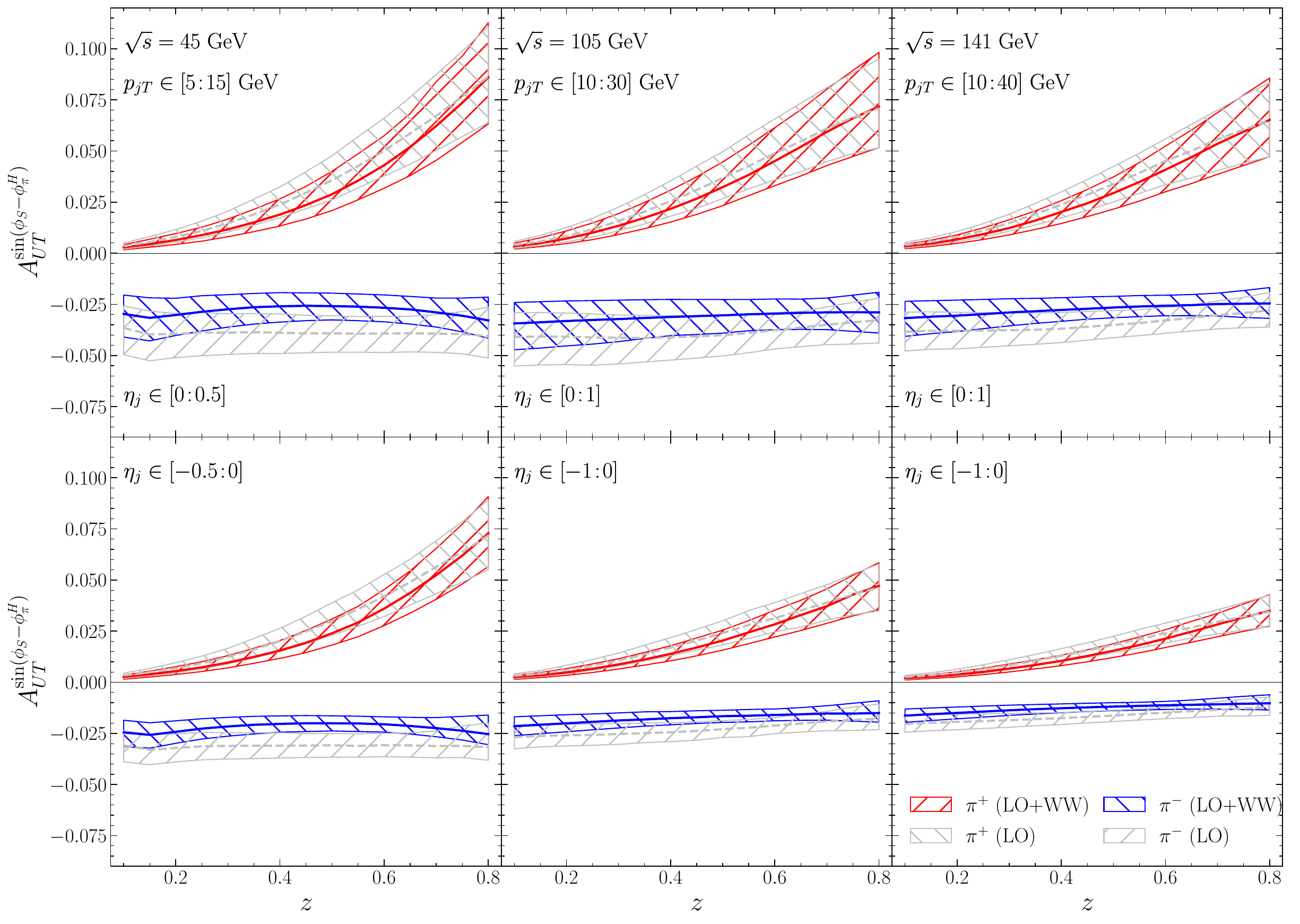}
\caption{Collins asymmetries for for $\ell p^\uparrow\to {\rm jet}\,\pi^\pm\,X$ as a function of $z$: LO (dashed lines and gray bands) and LO+WW (solid lines and colored bands). Left to right: 45, 105, 141 GeV, with the corresponding $p_{{\rm j}T}$ ranges, see legend. Upper panels: forward rapidities; lower panels: backward rapidities. Uncertainty bands at $2\sigma$ CL.}
\label{fig:AUT-vs-z}
\end{figure*}

Different and more interesting features appear in the $z$ dependence as shown in Fig.~\ref{fig:AUT-vs-z}. More precisely, while for $\pi^-$ production we find an almost flat behavior with a size around 2-3\% in the backward region and 3\% in the forward region, the asymmetries for $\pi^+$ production increase as a function of $z$ reaching values around 8\% in the forward region and being still around 3-6\% also for negative pseudorapidities values, depending on the $\sqrt{s}$ value.

The reasons for this behavior are the following:
$i)$ as discussed above, in the forward region the quark‑initiated channel largely dominates (at lower energies this is also the case in the backward region). This explains the differences between the forward and backward regions at intermediate and large energies, where gluon‑initiated contributions start playing a role;
$ii)$ the (favored) unfavored Collins fragmentation functions obtained from the fit of Ref.~\cite{Boglione:2024dal} exhibit, at large $z$, an (almost) identical behavior to that of the corresponding unpolarized fragmentation functions, with the unfavored ones being suppressed;
$iii)$ the process is dominated by the up‑quark contribution, reflecting the combined effect of its large transversity distribution and the associated large electric charge factor.

This implies that at large $z$ for $\pi^+$ production the leading term in the numerator is given by the convolution of the transversity for the up quark with the favored Collins FF (both large), while for $\pi^-$ the two terms entering the numerator, $4h_{1u}\otimes H_1^{\perp\rm{unf}} + h_{1d}\otimes H_1^{\perp\rm{fav}}$, involve always one suppressed factor. On the contrary, at low-$z$ values ($z \lesssim 0.2$) the suppression of the favored Collins FF (see Eq.~(\ref{eq:Collins-NC})) leads to almost negligible $\pi^+$ azimuthal asymmetries. These expectations could be directly tested at the EIC.

This also explains the difference with respect to the corresponding azimuthal asymmetries in $pp$ processes, where gluon contributions to the denominator are non-negligible in all kinematic regions and, at the same time, no analogous up-quark dominance (driven by the charge factor) emerges. This combined effect, while still showing an increasing behavior with $z$ (see Fig.~1 of Ref.~\cite{DAlesio:2025jmr}) leads to much smaller absolute values (less than 2\%).

\begin{figure*}[t]
\centering
\includegraphics[width=5.7cm,keepaspectratio]{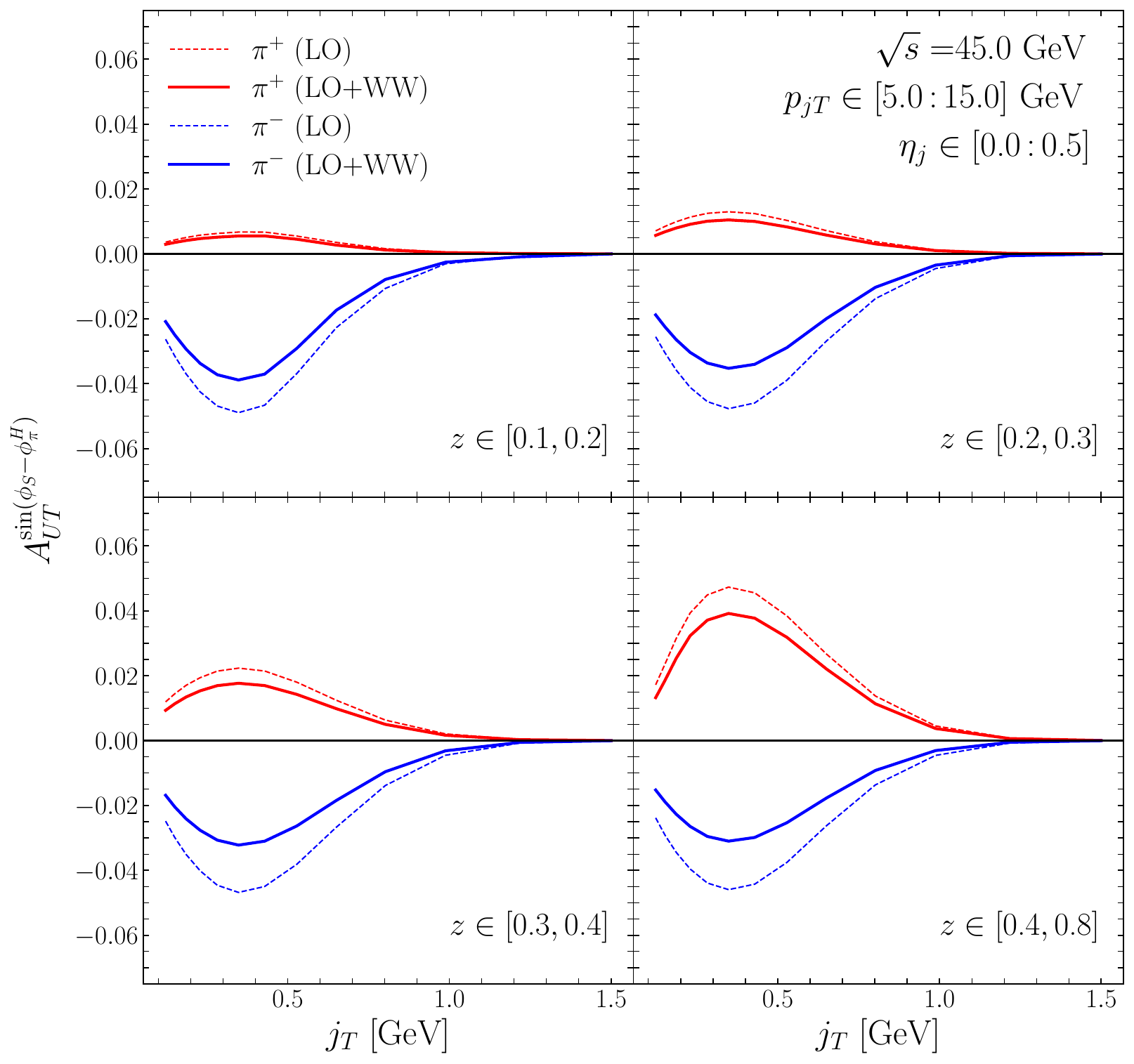}
\includegraphics[width=5.7cm,keepaspectratio]{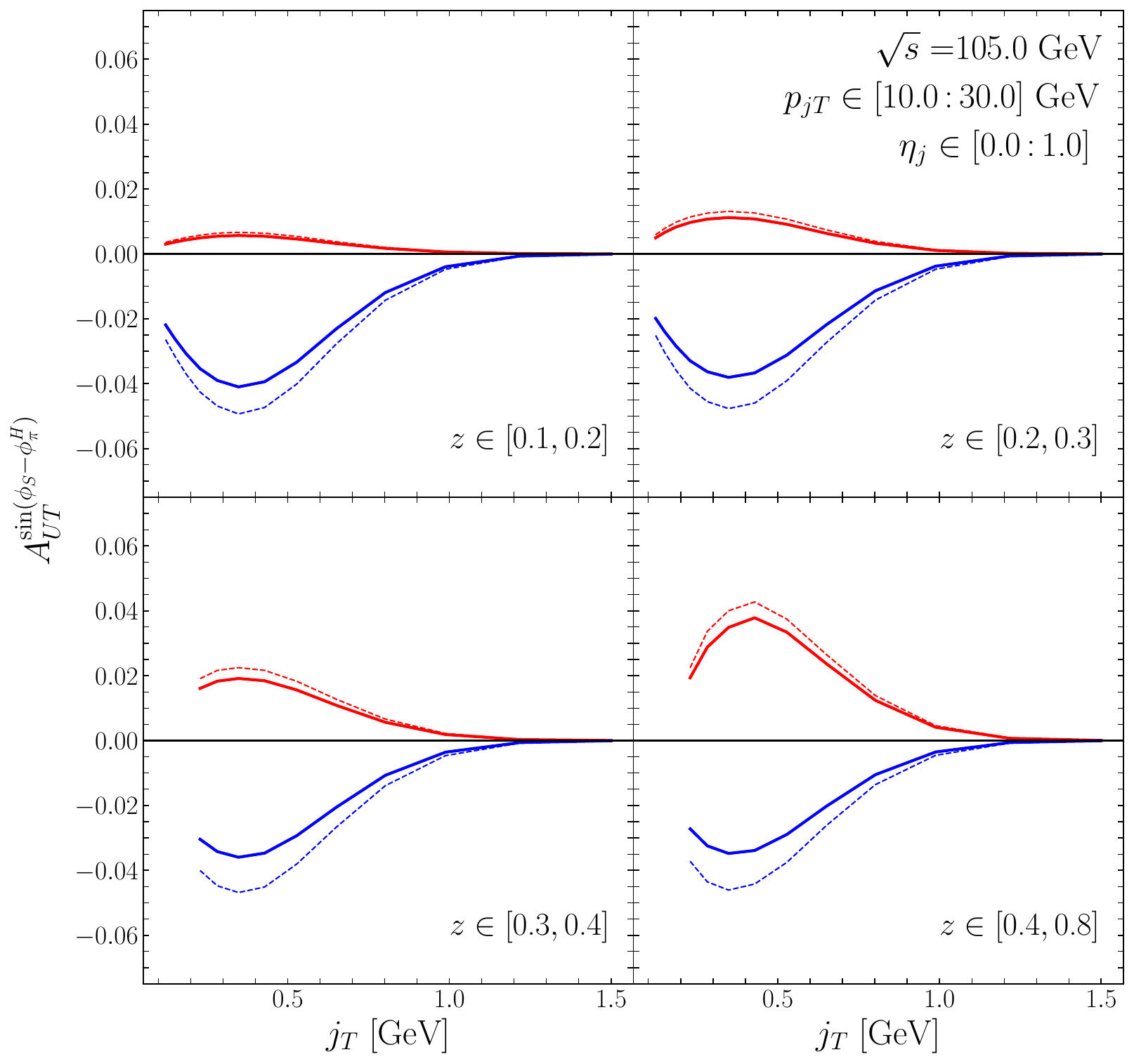}
\includegraphics[width=5.7cm,keepaspectratio]{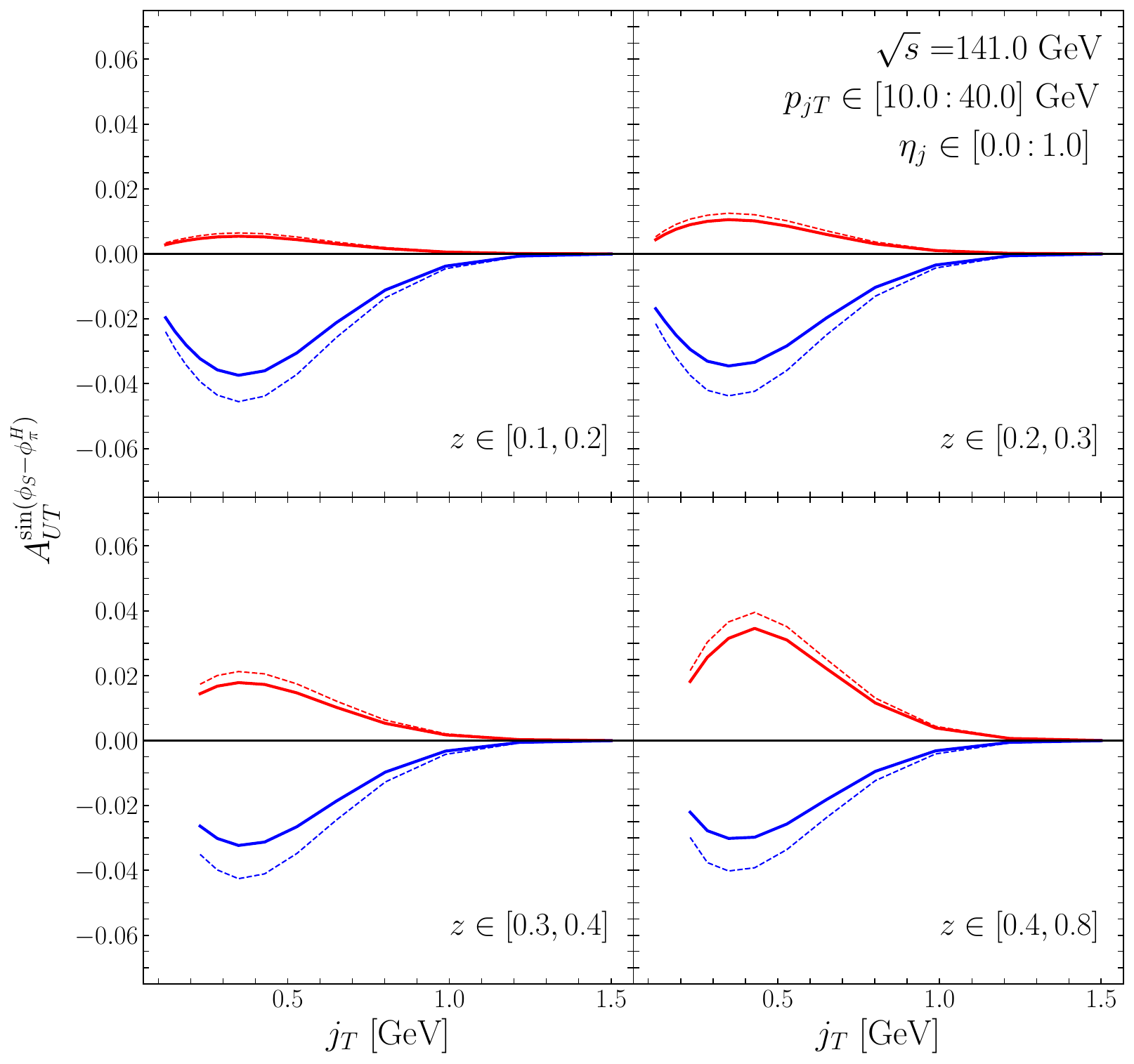}\\
\includegraphics[width=5.7cm,keepaspectratio]{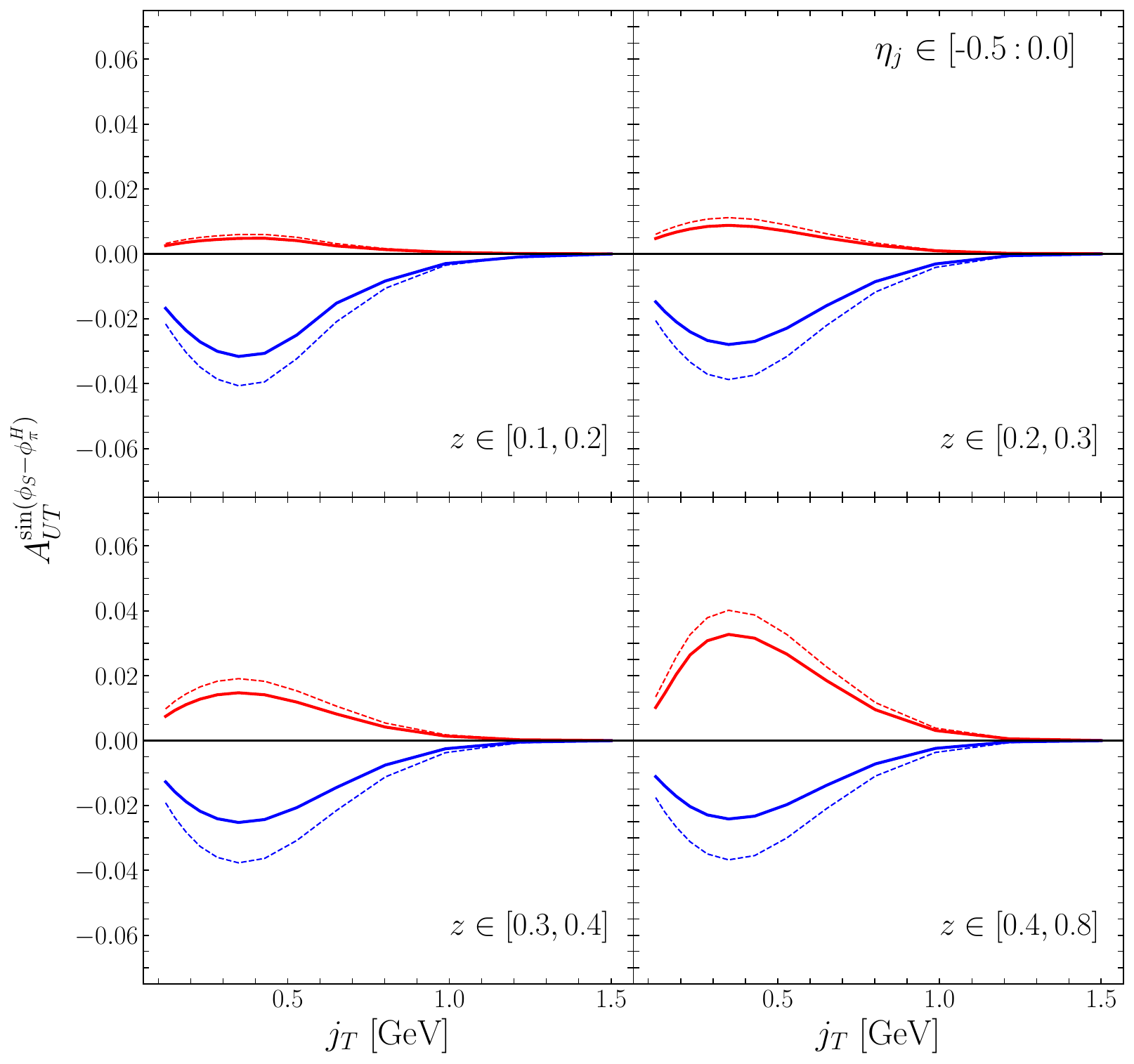}
\includegraphics[width=5.7cm,keepaspectratio]{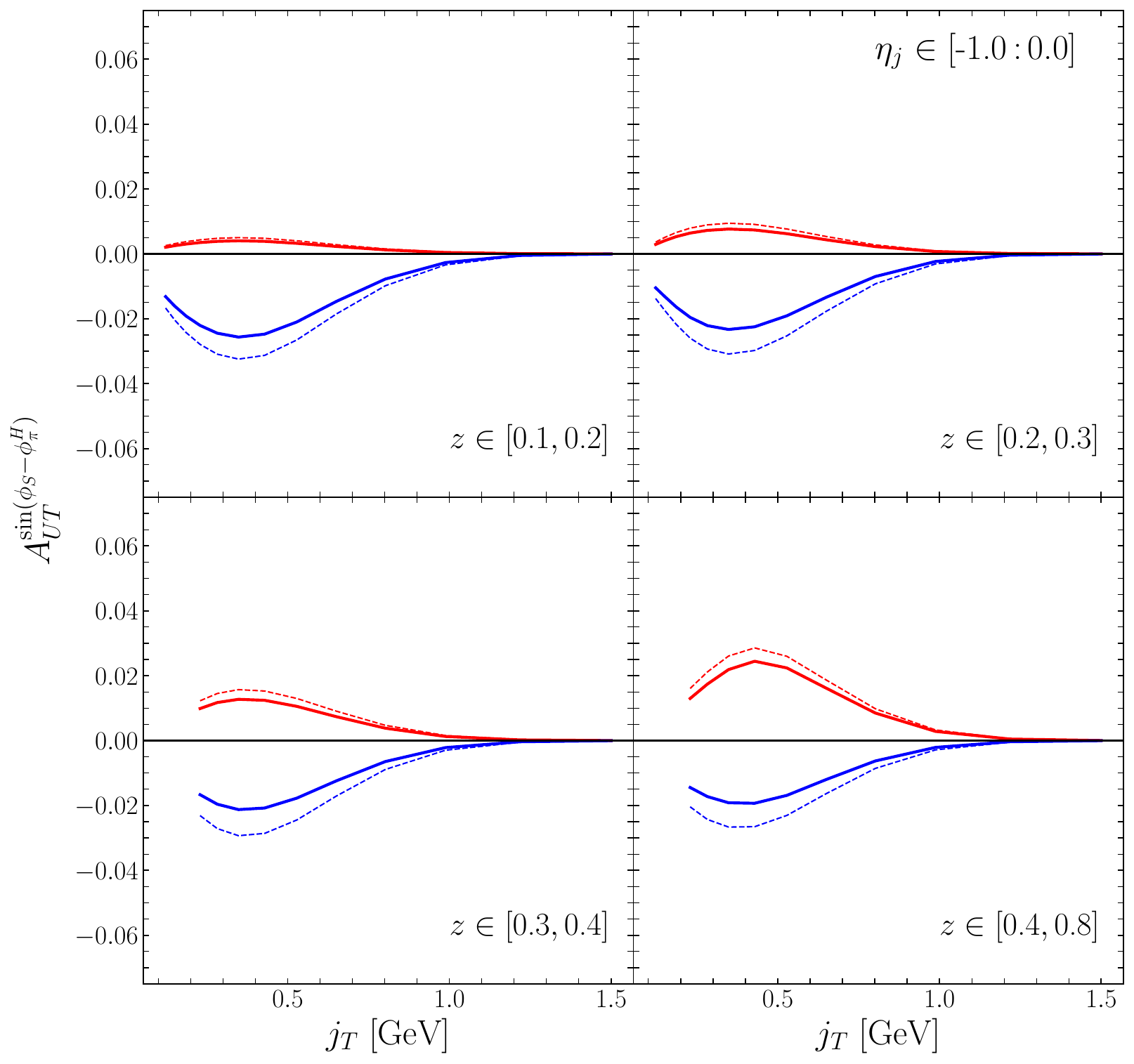}
\includegraphics[width=5.7cm,keepaspectratio]{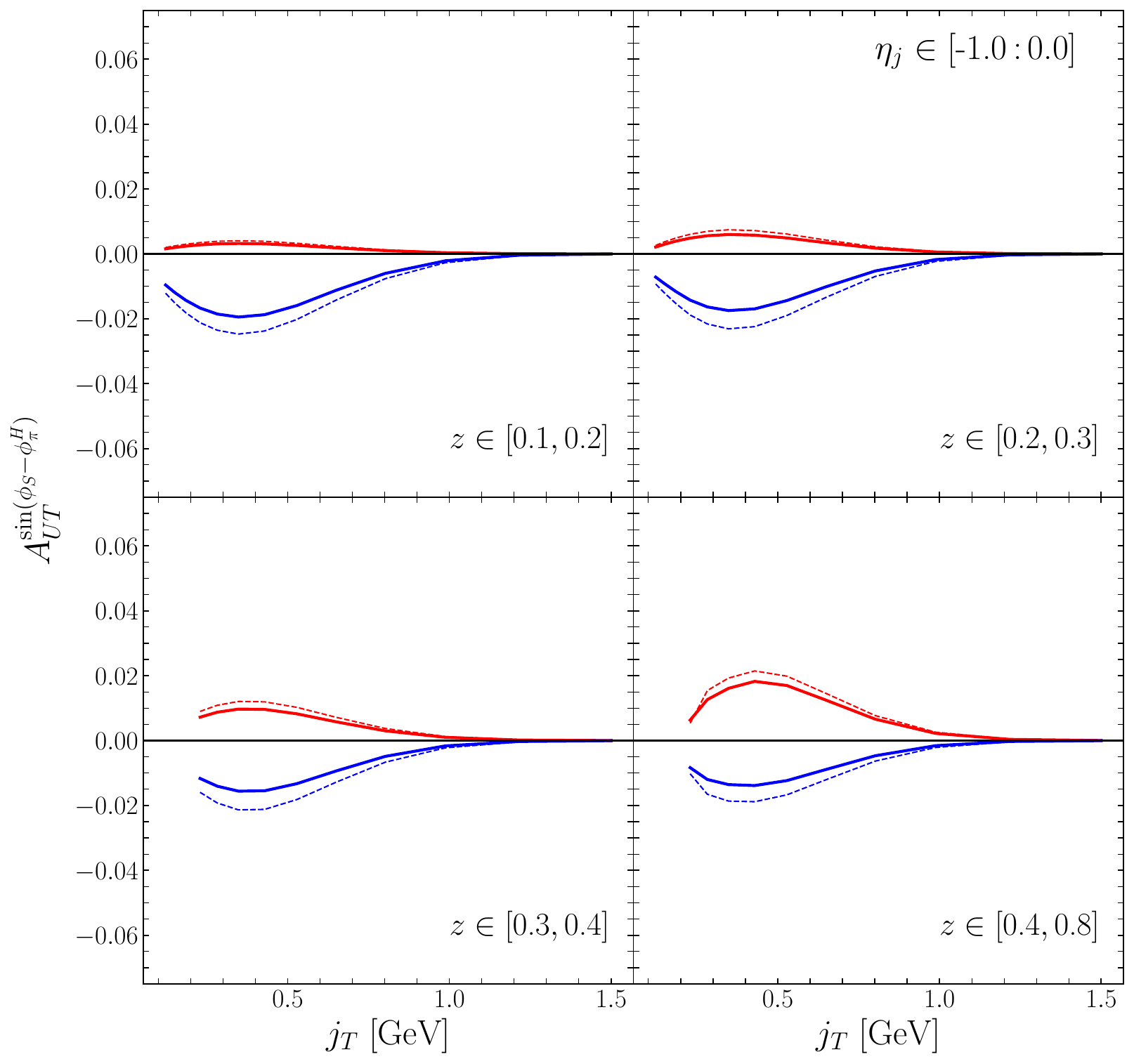}
\caption{Collins asymmetries for $\ell p^\uparrow\to {\rm jet}\,\pi^\pm\,X$ as a function of $j_T$ for different $z$ bins, only central values. Left to right: $\sqrt s=$ 45, 105, 141 GeV, with the corresponding $p_{jT}$ ranges, see legend. Upper panels: forward rapidities; lower panels: backward rapidities.}
\label{fig:AUT-vs-jT}
\end{figure*}

Lastly, in Fig.~\ref{fig:AUT-vs-jT}, we consider the $j_T$ dependence of the Collins asymmetries for different $z$ bins, where once again we find similar features.
More precisely we show the predictions for three energies, 45, 105, 141 GeV, left to right panels and for the forward (upper panels) and backward (lower panels) rapidities.
It is interesting to observe that in some cases the azimuthal asymmetry could be sizable.
Namely, for $\pi^+$ this happens in the intermediate $z$ bins (recall the suppression of the favored Collins FF at small $z$), while for $\pi^-$ this features is generally preserved for all $z$ bins (unsuppressed unfavored Collins FF coupled to the large transversity for up quarks).
This different behavior could be explored at the EIC, providing new insights on the flavor separation of the Collins FFs, together with an improvement in their parametrizations.

That said, a word of caution is warranted. Indeed, one has to keep in mind that these estimates are based on a very simple form of the intrinsic transverse momentum dependence of the Collins FFs, a Gaussian-like one.
This is one of the reasons for showing only the central values in different $z$ bins, without the corresponding uncertainty bands, focusing primarily on the $z$-$j_T$ correlations.
More importantly, a more flexible parametrization, including $z$ and flavor dependencies in the Gaussian widths, or a proper implementation of TMD jet FFs would give a  more solid picture.
Further detailed investigation is left to future work.
Overall, this suggests that this comparatively simple process may provide valuable insight into these still open issues.

In summary, this analysis, focused on Collins azimuthal asymmetries in $\ell p$  processes, highlights several interesting features. On the one hand, it allows for a test of the TMD factorization hypothesis in a comparatively simple process, as well as of the universality properties of the Collins fragmentation functions. On the other hand, it provides the opportunity to significantly improve our current knowledge of the Collins FF, including its flavor decomposition and intrinsic transverse‑momentum dependence, as well as of the transversity distribution in the large‑$x$ region. Finally, since this process is largely dominated by quark‑initiated channels, the backward region offers an ideal probe to access the still poorly known sea‑quark transversity.

\section{Conclusions}\label{sec:conclusions}
In the present work we have extended a previous phenomenological study of the Collins asymmetries for hadron-in-jet production in $pp$ collisions to the complementary $\ell p$ process, employing a scheme where TMD effects are restricted only to the fragmentation mechanism.
Encouraged by the quite good phenomenological agreement with experimental $pp$ data we have computed new predictions for the EIC kinematics, still employing the transversity and Collins functions as extracted from SIDIS and $e^+e^-$ annihilation processes.
Our estimates have been obtained going beyond the LO accuracy, by including the contribution from quasireal photon exchange, that, as shown, could play a relevant role.

The main goal of this study (combined with the corresponding one for $pp$ collisions) is to explore the universality issue, together with the TMD factorization hypothesis in non standard TMD processes.
To this aim we have considered different kinematical regions, in rapidities and transverse momenta of the jet and the observed hadron, highlighting the potential role of this analysis.

Another important outcome worth being stressed is that, despite new partonic channels opened by the WW contribution, like those initiated by gluons, the dominant role played by quark-initiated partonic channels is preserved. This provides a much clearer access to the transversity distribution, allowing at the same time to extract information on its sea-quark component, not directly achievable in other processes. In this respect the study of the Collins asymmetries in this process represents a formidable tool to test the overall framework and improve our knowledge on several fundamental issues.

Further studies can definitely help to get a more robust picture. The main issues to be explored are certainly the use of a more flexible parametrization of the Collins FFs, including the $z$ and flavor dependence in the Gaussian widths and/or, even more important, a more accurate treatment of the jet fragmentation mechanism, beyond the LO accuracy.
This will be carried out in a future study where we plan to reanalyze the Collins azimuthal asymmetries for hadron-in-jet production both in $pp$ and $\ell p$ collisions.

\section*{Acknowledgments}

The work of U.D.~is partially supported by Fondazione di Sardegna under the projects ``Journey to the center of the proton'', No.~F23C25000150007 (University of Cagliari). C.F.~is supported by the European Union’s Horizon Europe research and innovation programme under the Marie Skłodowska-Curie grant agreement n.~101150792 (STAT-TMDs). The work of M.Z. was partially supported by the U.S. Department of Energy contract No.~DE-AC05-06OR23177, under which Jefferson Science Associates, LLC operates Jefferson Lab, and conducted in part under the Laboratory Directed Research and Development Program at
Thomas Jefferson National Accelerator Facility for the U.S. Department of Energy.

\bibliographystyle{elsarticle-num}
\bibliography{references}

\end{document}